\documentclass[
aps,
prb,
reprint,
letterpaper,
superscriptaddress,
nofootinbib
]{revtex4-2}

\usepackage{amsmath,amssymb,bm,mathtools}
\usepackage{graphicx,xcolor,booktabs}
\usepackage[colorlinks=true,citecolor=blue!55!black,linkcolor=blue!55!black,urlcolor=blue!55!black]{hyperref}
\DeclareUnicodeCharacter{2212}{\ensuremath{-}}
\DeclareUnicodeCharacter{2009}{\,}
\DeclareUnicodeCharacter{00A0}{~}
\DeclareUnicodeCharacter{00BD}{\ensuremath{\tfrac{1}{2}}}
\DeclareUnicodeCharacter{015E}{\c{S}}

\newcommand{\ii}{\mathrm{i}}
\newcommand{\ee}{\mathrm{e}}
\newcommand{\dd}{\mathrm{d}}
\newcommand{\tup}{t_{\uparrow}}
\newcommand{\tdown}{t_{\downarrow}}

\begin{document}

\title{Non-Hermitian Two-Kink Control of Corner States in Twisted Chern Bilayers}

\author{Xiao-Ming Zhao}
\email{xmzhao@ustb.edu.cn}
\affiliation{Department of Physics and Institute of Theoretical Physics, University of Science and Technology Beijing, Beijing 100083, China}
\affiliation{{Key Laboratory of Multiscale Spin Physics (Ministry of Education), Beijing Normal University, Beijing 100875, China}}
\author{Su-Peng Kou}
\email{spkou@bnu.edu.cn}
\affiliation{{Key Laboratory of Multiscale Spin Physics (Ministry of Education), Beijing Normal University, Beijing 100875, China}}
\affiliation{ {Center for Advanced Quantum Studies, School of Physics and Astronomy, Beijing Normal University, Beijing 100875, China}}

\begin{abstract}
Corner states are bound by boundary-mass defects, but this geometric anchoring restricts their spatial and internal control. Focusing on twisted bilayers assembled from Chern insulators with opposite Chern numbers, we show that a mass-phase kink carried by a non-Hermitian interlayer mass and an independent imaginary-momentum kink separate the spectral, in-plane, and interlayer control of mass-defect corner states. The mass-phase kink creates the parent corner pole and fixes its energy and intrinsic decay, while directed interlayer tunneling encodes its layer polarization. The imaginary-momentum texture instead reshapes the closed-boundary envelope and redistributes stationary right-state weight between corners. A twisted Qi--Wu--Zhang bilayer verifies this division on arbitrary, including concave, boundaries. A $30^\circ$ twisted Haldane dodecagon realizes twelve finite-energy parent states whose in-gap spectral manifold persists while boundary patterns redistribute its normalized right-subspace weight toward six, three, two, or one target corners. The two kinks therefore cooperate or compete while keeping spectral ancestry, spatial weight, and layer character distinct, thereby establishing a unified framework for controlling mass-defect corner states across disparate boundary geometries and mass textures.
\end{abstract}

\maketitle

\textit{\textbf{Introduction.}---} Topological phases organize robust boundary transport, while higher-order topology extends this hierarchy to hinges and corners \cite{BBHScience,Langbehn,SchindlerSciAdv}. Two-dimensional corner states can arise through distinct mechanisms, including quantized multipole moments and Wannier centers \cite{BBHScience,BBHPRB}, crystalline-symmetry-constrained boundary masses \cite{Langbehn,Geier2018,SongFang,Khalaf}, obstructed or boundary-obstructed limits \cite{KhalafBOTP,Trifunovic2019,vanMiertOrtix}, breathing-lattice dimerization \cite{Ezawa}, and externally gapped first-order topological edges \cite{RenQiao}; a broader defect classification is given in Ref.~\cite{TeoKane}. Electronic and mechanical realizations are reported in Refs.~\cite{SchindlerNatPhys,SerraGarcia,Peterson}, while circuits, acoustic media, and photonic platforms provide complementary demonstrations \cite{Imhof,Xue,Ni,Fan,ZhangAcoustic}. These mechanisms are not equivalent. Here we focus on the mass-defect class in which one spectrally isolated pair of counterpropagating boundary modes is gapped by an orientation-dependent mass.

Coupling two topological layers with opposite edge helicities or opposite Chern numbers provides a direct realization of this boundary construction \cite{LiuCoupledTI,Miao2025}. Their layer-resolved edge modes counterpropagate, while interlayer tunneling projects to a mass whose sign or complex phase varies with edge orientation and stacking registry. Twist therefore converts geometry into a boundary-mass texture, producing continuously tunable $\pi$ kinks in twisted QWZ bilayers and symmetry-quantized fractional kinks in $30^\circ$ Haldane-type dodecagons \cite{Miao2025,Spurrier2020}. The mass texture determines the spectral parent state, but geometric anchoring alone offers limited control over where its wave-function weight is concentrated.

Non-Hermitian band theory provides a distinct localization mechanism. Complex spectra and point gaps organize its band topology \cite{BenderBoettcher,RudnerLevitov,GongPRX,KawabataClass}, while non-Bloch bulk--boundary correspondence and the skin effect reshape open-boundary modes \cite{ShenZhenFu,YaoWang,YaoSongWang,YokomizoMurakami,LeeThomale,Okuma}. Their interplay with higher-order topology produces gain/loss-induced corner states \cite{LuoZhang,LiGainLoss}, hybrid skin--topological modes \cite{Lee2019,ZhuGong}, and higher-order skin accumulation \cite{KawabataHigherSkin,FuNested}. Related boundary and corner localization has been observed in circuits \cite{Helbig,Xiao}, acoustic media \cite{Zou2021,ZhangHigherSkinExp}, and photonic systems \cite{Wang2022,Tang2023,Zheng2024}. Yet a bright non-Hermitian corner can represent either a spectrally inherited defect state or skin-accumulated boundary weight, so its spectral origin, in-plane visibility, and internal composition must be diagnosed separately.

\begin{figure*}[t]
\centering
\includegraphics[width=0.95\textwidth]{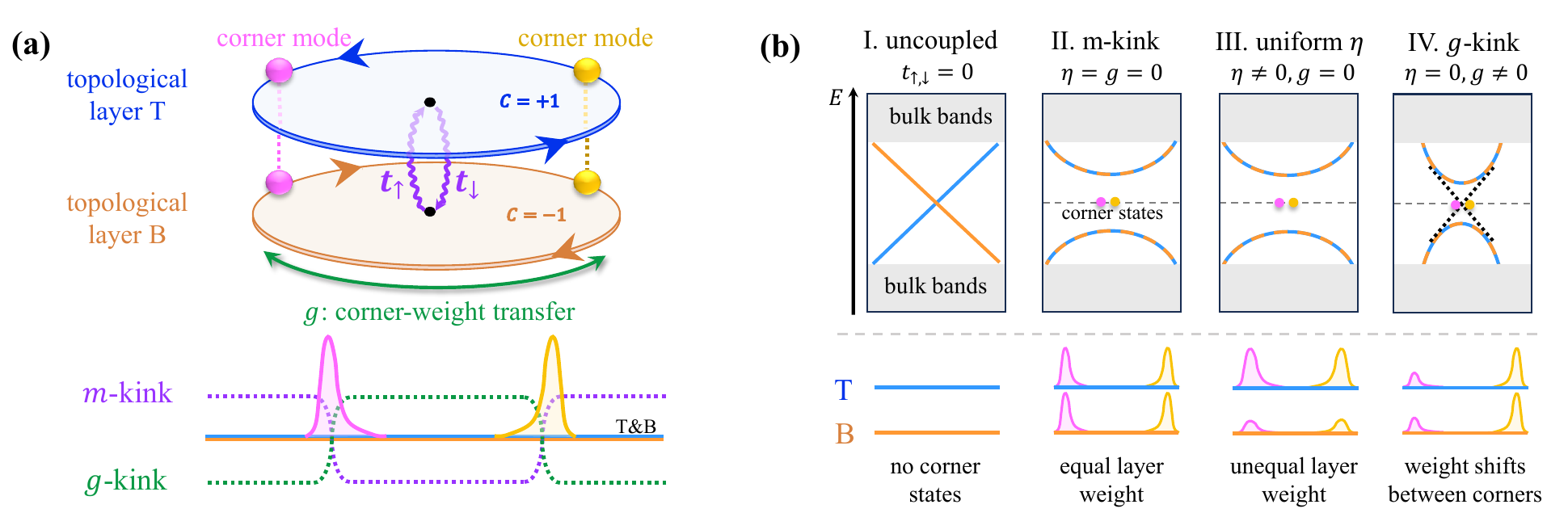}
\caption{\label{fig:framework}\textbf{Two-kink control in twisted Chern bilayers.} (a) Top: opposite-chirality Chern layers are coupled by $t_{\uparrow}\ne t_{\downarrow}$ and carry layer-opposite imaginary boundary potentials; the spheres denote two-layer corner modes. Bottom: the projected counterpropagating edge channels contain two mass-phase kinks (purple, $m$-kinks) and two independently generated imaginary-momentum kinks (green, $g$-kinks). (b) Starting from uncoupled channels (I), reciprocal tunneling opens an edge gap and creates layer-balanced parent states at the mass kinks (II). Uniform $\eta$ polarizes their layer weights (III), whereas a $g$-kink transfers normalized right-state weight between parent corners without adding eigenstates (IV).}
\end{figure*}

Within this mass-defect class, we formulate corner-state control as the cooperation and competition of two boundary kinks. A mass-phase kink carried by a non-Hermitian interlayer mass creates the parent pole, whereas the directed-amplitude imbalance of the same mass encodes its layer spinor. In the two realizations studied here this imbalance is edge independent, $\Delta\eta=0$, and therefore polarizes the parent state without adding a mass-modulus defect. An independent imaginary-momentum kink changes the physical tail rates and globally redistributes normalized right-state weight around a closed boundary. Reservoir engineering and cascaded coupling provide routes to directed hopping \cite{Gardiner,Carmichael,MetelmannClerk,ClerkRMP}, while layer-opposite boundary gain and loss projects to imaginary momentum along chiral edges \cite{Liu2024,Ma2024}. We test this separation in complementary twisted Chern bilayers: QWZ supplies a continuously varying real mass on arbitrary straight edges \cite{QiWuZhang,Miao2025}, whereas the $30^\circ$ Haldane dodecagon supplies a symmetry-quantized complex-mass clock \cite{Haldane,Spurrier2020}. Together they establish a unified boundary control law in which the mass-phase kink selects the spectral parent subspace, the imaginary-momentum texture selects its in-plane weight, and interlayer nonreciprocity selects its layer polarization.

\textit{\textbf{Projected boundary theory.}---}
Figure~\ref{fig:framework} summarizes the mass-defect construction considered here: twist generates the parent mass-phase texture, directed interlayer tunneling fixes the layer spinor, and a separately imposed imaginary-momentum texture redistributes the right-state envelope along the closed boundary. On a straight edge $j$, projection of the microscopic bilayer Hamiltonian onto one counterpropagating layer-edge pair gives \cite{LiuCoupledTI,Miao2025,Spurrier2020}
\begin{equation}
H_j=-\ii v_j\rho_z(\partial_s-g_j)
+m_j\begin{pmatrix}0&\ee^{\eta-\ii\phi_j}\\
\ee^{-\eta+\ii\phi_j}&0\end{pmatrix}.
\label{eq:edge}
\end{equation}
Here $m_j>0$ and $\phi_j$ are the magnitude and phase of the reciprocal projected mass, $g_j$ is the projected imaginary momentum, and $\rho_i$ act in the two edge-channel space. For real-positive directed tunneling, $\eta=\tfrac12\ln(\tup/\tdown)$. Equation~\eqref{eq:edge} follows from two directed matrix elements of a distance-decaying interlayer kernel: their product fixes $m_j^2$, and their ratio fixes $\eta$. The microscopic projection and continuous edge gauge are derived in Supplemental Material (SM), Ref.~\cite{SM}, Sec.~S1; Lindblad and cascaded-coupling realizations are given in Sec.~S2.

The second term of Eq.~\eqref{eq:edge} is a non-Hermitian mass matrix. Across a corner, $\Delta\phi$ is its phase kink and $\Delta\eta$ would be its modulus-asymmetry kink. We set $\Delta\eta=0$ in the main text: uniform $\eta$ changes the layer composition but not the matching determinant. The only mass defect in the QWZ and Haldane validations is therefore the projected phase kink. The more general $\Delta\eta\ne0$ matching problem is stated in SM Sec.~S7.

Continue the edge basis through a corner and unfold the adjoining edges I and II into two rays. For equal reciprocal masses and velocities, spinor matching at a mass-phase step $\Delta\phi=\phi_{\rm II}-\phi_{\rm I}$ yields
\begin{equation}
E_c=m\cos(\Delta\phi/2).
\label{eq:parent}
\end{equation}
The same matching gives $\kappa=(m/v)\sin(\Delta\phi/2)$ on the oriented branch $0<\Delta\phi<2\pi$ \cite{JackiwRebbi,GoldstoneWilczek,Spurrier2020}. Thus a $\pi$ kink is a zero mode with $\kappa=m/v$, whereas the chosen $\pi/2$ branch has $E_c=m/\sqrt2$ and $\kappa=m/(\sqrt2v)$. Unequal masses and velocities obey the determinant derived in SM Sec.~S3.

The order of this construction matters. The determinant that produces Eq.~\eqref{eq:parent} is first solved in the reciprocal frame, so it decides whether a defect pole exists and where it lies relative to the edge continuum. Only after that pole is fixed is the imaginary momentum restored to determine the observable tails. This separates a spectral statement---a zero of the matching determinant---from a localization statement about the corresponding right eigenvector. It also prevents a skin maximum at an otherwise smooth mass junction from being misidentified as an independently created topological state.

Restoring the imaginary momentum multiplies the reciprocal-frame solution by an exponential gauge factor. With coordinates directed away from the corner along each ray, the physical inverse decay rates are
\begin{equation}
\lambda_{\rm I}=\kappa_{\rm I}+g_{\rm I},\qquad \lambda_{\rm II}=\kappa_{\rm II}-g_{\rm II}.
\label{eq:lambda}
\end{equation}
For an isolated two-ray corner, a matched parent is square integrable if and only if both real parts are positive. These inequalities test the physical tails after mass matching; they cannot create a pole at a nonparent corner. Increasing both rates makes the two kinks cooperate, whereas crossing either zero-rate line makes imaginary momentum compete with and release one parent tail.

\begin{figure}[!t]
\centering
\includegraphics[width=\columnwidth]{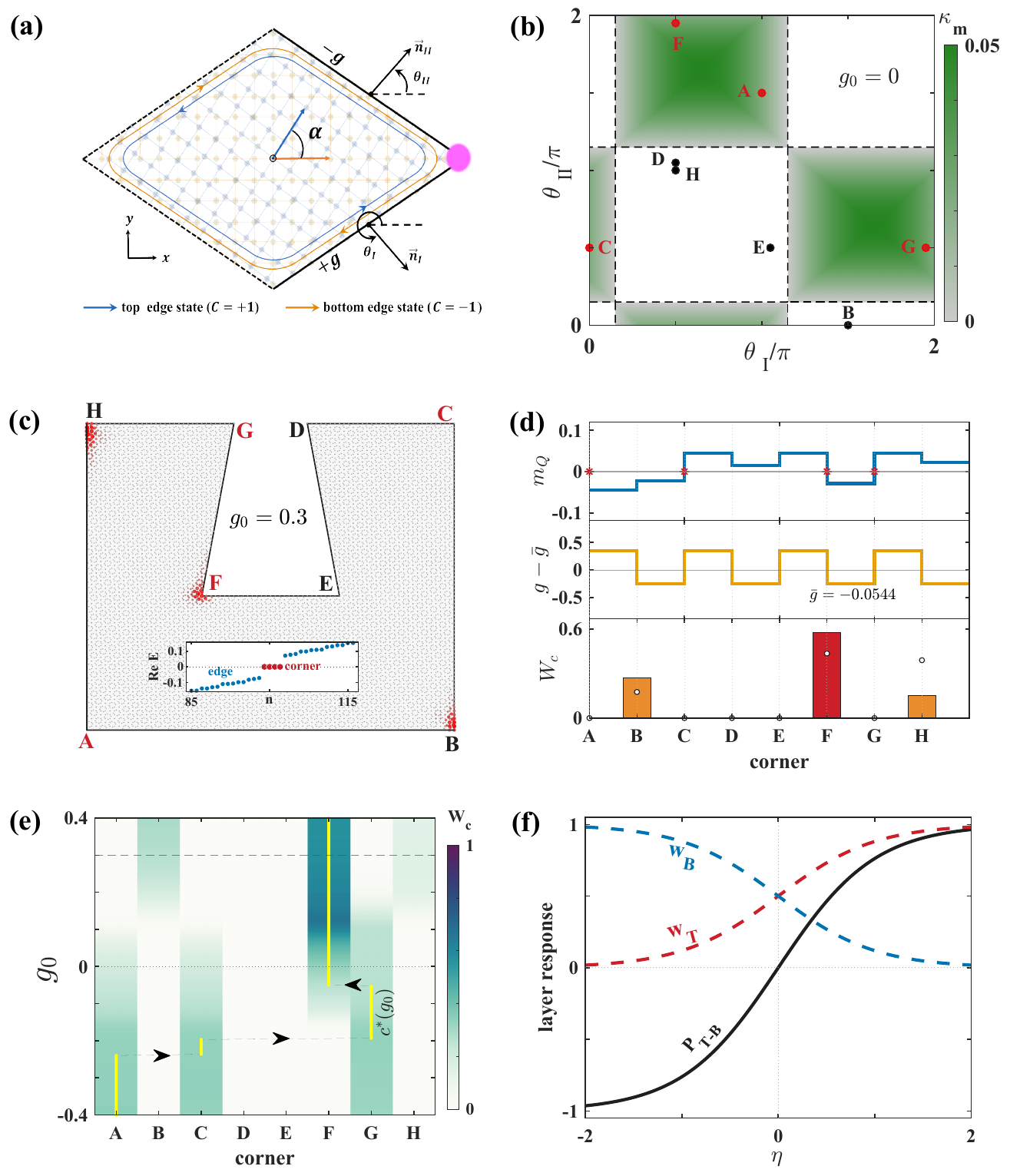}
\caption{\label{fig:qwz}\textbf{Arbitrary-edge QWZ verification.} (a) Twisted bilayer and adjacent straight-edge normals $\theta_{\rm I,II}$. (b) At $g_0=0$, color gives $\kappa_m$ where $m_Q(\theta_{\rm I})m_Q(\theta_{\rm II})<0$; A--H label the vertices in (c). (c) Concave-octagon density and real spectrum: A, C, F, and G are the $g_0=0$ parents, whereas an alternating $\pm g_0$ shell at $g_0=0.3$ concentrates the four tracked near-zero modes (red inset) mainly at B, F, and H. (d) Perimeter profiles of $m_Q$, $g-\bar g$, and corner weight along A$\to$H$\to$A. (e) Global-envelope weights versus $g_0$; the light trajectory marks the brightest sector and the red dashed line the value in (c). (f) Layer weights and polarization versus $\eta$ at fixed $\sqrt{\tup\tdown}=0.05$. Parameters are given in SM Sec.~S5.}
\end{figure}

An isolated-wedge criterion is incomplete on a finite polygon. For perimeter length $L_\partial$, let $\bar g=L_\partial^{-1}\oint g(s)\,\dd s$ and define
\begin{equation}
\Phi(s)=\int_0^s[g(s')-\bar g]\,\dd s'.
\label{eq:closed}
\end{equation}
Because $\Phi(L_\partial)=\Phi(0)$, the transformation is single valued and the right-state density satisfies
\begin{equation}
\rho_g(s)\propto \ee^{2\Phi(s)}\rho_{\bar g}(s).
\label{eq:envelope}
\end{equation}
The same transformation gives $H[g]=\ee^{\Phi}H[\bar g]\ee^{-\Phi}$ \cite{HatanoNelson,Liu2024,Ma2024}. Hence a zero-mean texture is exactly isospectral to $g=0$ in the ideal scalar single-channel loop, whereas nonzero $\bar g$ is an imaginary flux that changes the monodromy and the boundary spectrum. The function $g-\bar g$ locates envelope extrema, while $\Phi_c-\Phi_d$ compares corners. Finite shell width, edge-dependent projection, and edge--bulk mixing break the exact similarity and account for the small spectral drift in the lattice calculations (SM Secs.~S4 and S7).

The signs of $\lambda_{\rm I,II}$ classify the two tails of one ideal corner but cannot rank vertices on a finite loop. That ranking requires the cumulative integral in Eq.~\eqref{eq:closed}. Combining parent matching with extrema of $\Phi$ gives enhanced and suppressed parents, nonparent skin peaks, and dark nonparents. Here visibility means the normalized spatial weight of a spectrally identified right-eigenstate subspace; it is not by itself a driven steady-state intensity or an unconditional Lindblad occupation. The distinction between stationary mode weights, driven response, and conditional no-jump evolution is discussed in SM Sec.~S7. Spectral ancestry is fixed before the density is interpreted, which separates parent corner states from chiral boundary skin states occupying the same vertex.

Uniform $\eta$ is removed from Eq.~\eqref{eq:edge} by $S_\eta=\exp(\eta\rho_z/2)$, a nonunitary similarity transformation that changes right-state weights without changing the ideal spectrum \cite{YaoWang,Kunst}. It therefore leaves Eqs.~\eqref{eq:parent} and \eqref{eq:envelope} unchanged in the single-channel model, while the normalized right-state layer polarization is
\begin{equation}
P_{T-B}=\frac{w_T-w_B}{w_T+w_B}=\tanh\eta.
\label{eq:polarization}
\end{equation}
The coordinate-space factor $\ee^{\Phi(s)}$ and internal factor $S_\eta$ commute. Position and layer polarity are therefore independently controllable at fixed parent mass texture. Equation~\eqref{eq:polarization} assumes equal layer amplitudes in the reciprocal-frame corner spinor; the general expression contains the corresponding reference weights. The factorization requires one spectrally isolated edge pair, a nonzero mass product, uniform real-positive directed tunneling, and scalar $g(s)$; complex hopping, edge-dependent $\eta$, and multichannel mixing are treated in SM Sec.~S7.

\textit{\textbf{Continuous mass texture.}---}
We first test the two-kink framework on a continuously tunable mass texture, where arbitrary edge orientations separate the geometric locations of parent corner states from the vertices selected by imaginary momentum. For opposite-chirality QWZ layers twisted by $\alpha$, the straight-edge projection gives the signed canonical mass
\begin{equation}
m_Q(\theta)=\Lambda_Q(\theta)\sqrt{\tup\tdown}
\sin\!\left(\theta-\frac{\alpha}{2}\right)
\label{eq:mq}
\end{equation}
with $\Lambda_Q(\theta)>0$. Its sign is the two-valued phase of Eq.~\eqref{eq:edge}; hence edges I and II carry a parent zero mode when $m_Q(\theta_{\rm I})m_Q(\theta_{\rm II})<0$. This condition and $\kappa_{Q,j}=|m_Q(\theta_j)|/v_Q$ follow from the arbitrary straight-edge modes and directed interlayer projection in SM Sec.~S5 \cite{QiWuZhang,Miao2025}.

Figure~\ref{fig:qwz} separates parent-state existence from finite-boundary visibility. Panel (b) identifies A, C, F, and G as the four $g_0=0$ parent corners. After an alternating layer-opposite $\pm g_0$ shell is applied, the four-dimensional near-zero subspace remains spectrally identifiable, but at $g_0=0.3$ its summed right-state density is concentrated mainly at B, F, and H [panel (c)]. The perimeter-resolved $m_Q$, $g-\bar g$, and corner weights in panel (d), together with the full $g_0$ scan in panel (e), show that the cumulative envelope $\Phi$ rather than $\min(\lambda_{\rm I},\lambda_{\rm II})$ alone ranks vertices on the finite concave boundary. Consequently, the bright nonparent vertices B and H collect skin-redistributed boundary or parent-tail weight; they do not represent additional zero-mode eigenvalues. Panel (f) independently verifies $P_{T-B}=\tanh\eta$ at fixed spectral mass product.

\begin{figure}[!t]
\centering
\includegraphics[width=\columnwidth]{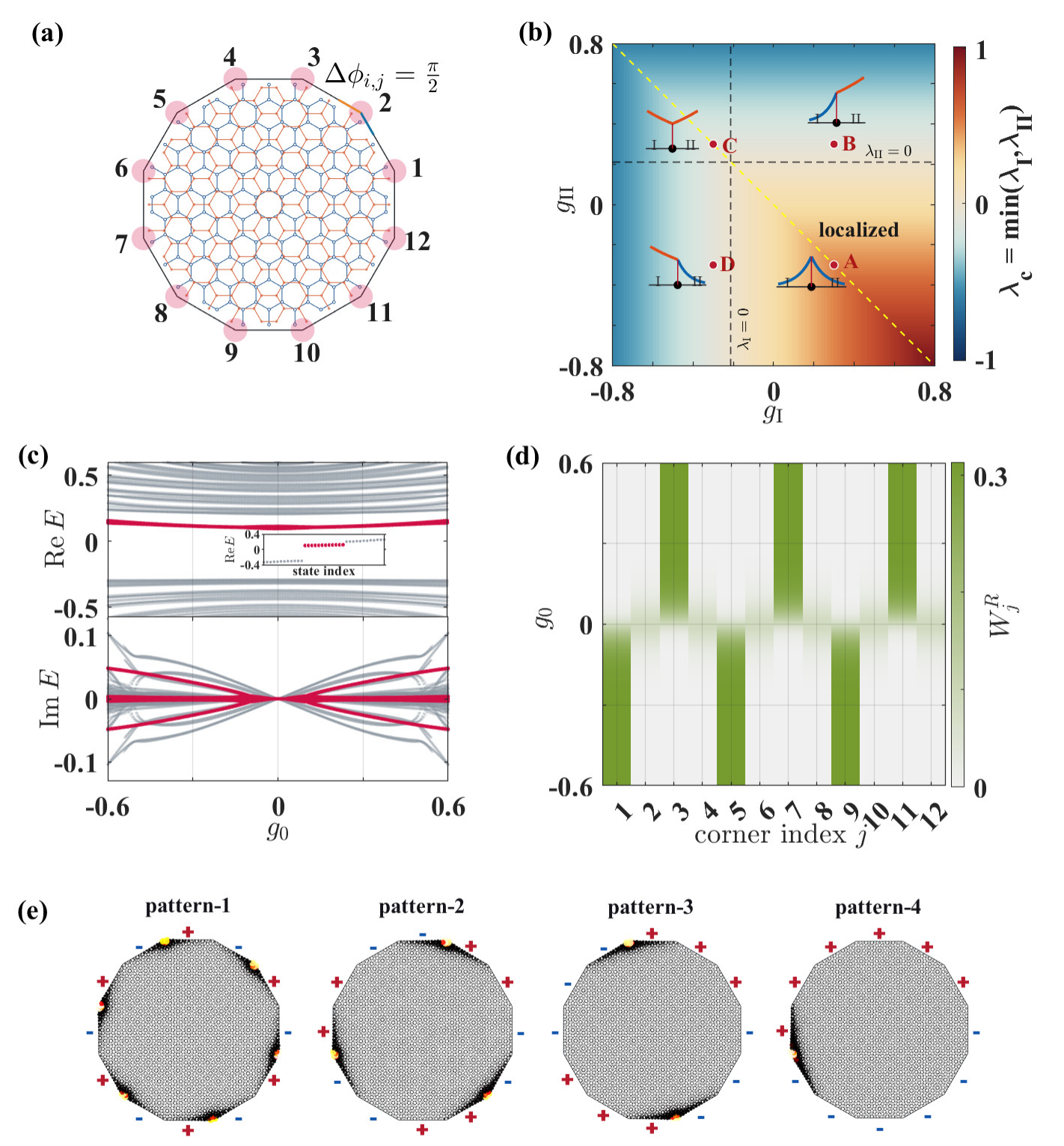}
\caption{\label{fig:haldane}
\textbf{Fractional-kink Haldane corner states.}
(a) A $30^\circ$ twisted dodecagon and its $C_{12}$ mass clock: $\Delta\phi=\pi/2$, $E_H=m_H/\sqrt2$, and $\kappa_H=m_H/(\sqrt2v_H)$. (b) Local two-ray map from $\lambda_{\rm I}=\kappa_H+g_{\rm I}$ and $\lambda_{\rm II}=\kappa_H-g_{\rm II}$. A--D denote the sign sectors $(++),(+-),(--),(-+)$; only A is locally normalizable. (c) Real and imaginary spectra of pattern-2 versus $g_0$. The twelve corner states (red) remain inside the real-energy edge gap; the inset at $g_0=0.3$ corresponds to the second density in (e). (d) Corner weights $W_j^R$ of this tracked subspace. (e) At $g_0=0.3$, pattern-1--4 aggregate the same twelve-state weight toward $6,3,2,$ and $1$ target corners. Here $m_H=0.30$, $v_H=1$, shell width $2.5$, and panels (c),(d) use $|g_0|\le0.6$. Displayed densities use $\eta=0$ to isolate $g$; uniform $\eta\ne0$ gives the Fig.~\ref{fig:qwz}(f) layer polarization. Parameters are given in SM Sec.~S6.}
\end{figure}

The concave geometry is more than a shape-specific illustration. Because $m_Q(\theta)$ is defined for every straight-edge normal, the same mass criterion treats convex and reentrant vertices without assigning a separate corner Hamiltonian to either type. At $g_0=0$, the four sign reversals identify the parent subspace before any non-Hermitian redistribution is applied. Turning on the boundary shell then provides a stringent test: a density maximum can move to a nonparent vertex even though the tracked near-zero subspace retains dimension four. The simultaneous perimeter plots of $m_Q$, $g-\bar g$, and corner weight expose precisely where a local two-ray prediction ceases to be a complete finite-system criterion.

\textit{\textbf{Fractional mass texture.}---}
We next examine a symmetry-fixed fractional mass texture, testing whether the same control framework extends to finite-energy parent states and the collective redistribution of their spatial weight. A regular dodecagonal cut of a $30^\circ$ twisted Haldane bilayer replaces the QWZ sign texture by a complex phase recurrence \cite{Haldane,Spurrier2020}. The registry contraction and the dressed twelvefold operation give
\begin{equation}
\phi_{j+1}-\phi_j=\frac{\pi}{2}.
\label{eq:haldane}
\end{equation}
Every geometric corner is therefore an oriented fractional parent mass-phase kink. For the phase clock in Eq.~\eqref{eq:haldane}, $E_H=m_H/\sqrt2$ and $\kappa_H=m_H/(\sqrt2v_H)$. The negative-energy solution belongs to the reversed clock, not to an additional branch at each corner (SM Sec.~S6).

The Haldane data are displayed at $\eta=0$ only to isolate the imaginary-momentum texture. A uniform $\eta\ne0$ applies the same internal similarity $S_\eta$ as in Eq.~\eqref{eq:polarization}, producing layer polarization without changing the $C_{12}$ phase recurrence or the ideal finite-loop envelope; this dependence is therefore not replotted. Figure~\ref{fig:haldane}(b) isolates the local cooperation and competition: the A--D sectors diagnose whether the two tails decay or grow away from an isolated mass kink. They do not rank vertices on the closed polygon; the accumulated envelope supplies that global information in panel (d). The twelve-dimensional parent manifold is identified at $g_0=0$ and continued by maximizing the principal-angle overlap of QR-orthonormalized right subspaces. Panel (c) shows that the tracked branches remain inside the real-energy boundary-state gap. Only after this spectral ancestry is fixed is the corner-resolved right-state weight evaluated, so co-localized boundary skin modes are not counted as extra corner eigenstates \cite{Lee2019,KawabataHigherSkin,ZhuGong,Ma2024}.

Panel (e) applies four commensurate sign patterns at $g_0=0.3$. The same twelve in-gap parent states aggregate toward six, three, two, and one target corners. Thus $12\rightarrow N$ denotes right-subspace weight aggregation, not eigenvalue loss or a quantized invariant. It remains meaningful while the edge gap and the separation between competing maxima of $\Phi$ stay finite (SM Sec.~S6).

\textit{\textbf{Summary.}---}
We have established a unified boundary description of two-kink control in twisted Chern bilayers. The mass-phase kink fixes the parent spectral sector, uniform interlayer nonreciprocity encodes its layer polarization, and imaginary momentum changes its local tails and finite-loop envelope. Their cooperation or competition can enhance or suppress a parent corner, or collect inherited weight at a nonparent vertex. The resulting bright-corner count is therefore distinct from the number of topological parent eigenstates.

The QWZ and Haldane results connect arbitrary boundary geometry and fractional mass textures through this same separation. Local decay describes individual tails, whereas the integrated envelope compares distant corners; in particular, Haldane weight aggregation retains the twelve-state parent manifold. While the parent sector remains identifiable within the edge gap, spatial redistribution and layer encoding can be tuned without redesigning the parent mass texture. This extends corner-state control beyond geometric anchoring by separating where a mode originates from where its stationary weight is concentrated and how it is distributed between layers. The unified framework thus provides a common design principle for independently tuning the spatial and internal character of both zero-energy and finite-energy mass-defect corner states.

\begin{acknowledgments}
This work is supported by Guangdong Basic and Applied Basic Research Foundation (Grant No. 2023A1515110081), Open Fund of Key Laboratory of Multiscale Spin Physics (Ministry of Education), Beijing Normal University (Grant No. SPIN2024K01), Fundamental Research Funds for the Central Universities (Grant No. FRF-TP-22-098A1, FRF-IDRY-24-28), National Key R\&D Program of China (Grant No. 2023YFA1406704), National Natural Science Foundation of China (Grant Nos. 12174030).
\end{acknowledgments}

\section*{Data availability}
The data and codes that support the findings of this study are available from the corresponding author upon reasonable request.

\end{document}